# Implementation of the KDamper concept using disc springs

Ioannis E Sapountzakis[1], Pavlos G Tranakidis[2] and Ioannis A Antoniadis[3]

[1,2,3]Mechanical Design and Control Systems Section,

Mechanical Engineering Department

National Technical University of Athens,

Heroon Polytechniou 9, 15780 Zografou, Greece

## ABSTRACT

The KDamper is a novel passive vibration isolation and damping concept, based essentially on the optimal combination of appropriate stiffness elements, which include a negative stiffness element. The KDamper concept does not require any reduction in the overall structural stiffness, thus overcoming the corresponding inherent disadvantage of the "Quazi Zero Stiffness" (QZS) or "Negative Stiffness" (NS) isolators, which require a drastic reduction of the structure load bearing capacity. Compared to the traditional Tuned Mass damper (TMD), the KDamper can achieve better isolation characteristics, without the need of additional heavy masses, as in the case of the TMD. Contrary to the TMD and its variants, such as the inerter, the KDamper substitutes the necessary high inertial forces of the added mass by the stiffness force of the negative stiffness element. Among others, this can provide comparative advantages in the very low frequency range. It should be mentioned, that the KDamping concept does not simply refer to a discrete vibration absorption device, but it consists a general vibration ab-sorption concept, applicable also for the design of advanced materials or complex structures. Such a concept thus presents the potential for numerous implementations in a large variety of technological applications, while further potential may emerge in a multi- physics environment. The paper proceeds to a review of the optimal design and selection of the parameters of the KDamper, which follows exactly the classical approach used for the design of the TMD. The negative stiffness elements have been replaced with a set of Belleville (disc) springs. It should be mentioned that two different cases for an initial displacement and an initial velocity have been considered. It is thus theoretically proven that the KDamper can inherently offer far better isolation and damping properties than the TMD. Finally, an application concerning the implementation of the KDamper for the design of a low frequency vertical vibration isolator is presented.

Corresponding author: Ioannis E Sapountzakis, PhD Candidate, Mechanical Design and Control System Section, Department of Mechanical Engineering, National Technical University of Athens, Heroon Polytechniou 9, 15780 Zografou, Greece, Email: isapountzakis@mail.ntua.gr

1. Introduction

The Tuned Mass Damper (TMD) has a long history, already more than 100 years. The TMD concept was first applied by [20]. A theory for the TMD was presented later in the paper by [34] . A detailed discussion of optimal tuning and damping parameters appears in [16]. Since then, numerous applications of various forms of TMDs have been reported. Some recent examples include vibration absorption in seismic or other forms of excitation of structures [15], wind and wave excitation in wind turbines by [8] and torsional vibrations in rotating and reciprocating machines by [31]. TMDs are available in various physical forms, including solids, liquids [8], or even active implementations [50]. The essential limitation of the TMD is that a large oscillating mass is required in order to achieve significant vibration reduction. Among others, this has prohibited the usage of TMDs in the automotive or aerospace sector.

In an attempt to reduce the requirements for heavy oscillating masses, the inerter concept has been introduced in early 2000s by [32]. The inerter is a two terminal element which has the property that the force generated at its ends is proportional to the relative acceleration of its terminals. This constant of proportionality is called "inertance" and is measured in kilograms. The main advantage of the inerter is that the inerter need not have large mass in order to achieve the same inertia effect as the additional mass of the TMD. However, according to [10] since inerters, dampers and springs can be connected in multiple configurations, the comparison of the structure of the Frequency Response functions of the inerter and of the TMD becomes very complicated. In 2005 the inerter was profitably used as a part of suspension in Formula 1 racing car under the name of "J-damper" [9]. Since then, other applications emerged, such as in suspensions of railway vehicles [47] or in seismic protection of structures [42]. Although the initial inerter configuration is for linear accelerations, rotary versions [28], or even active configurations [46] have been proposed. Still, to work efficiently, all considered devices have to be precisely tuned which can be hard to achieve or even impossible in some cases. Moreover, proposed TMDs with inerters suffer from susceptibility to detuning. Although variable inertance mechanical configurations have been proposed for this purpose [5], the essential limitation of the inerter is the complex and elaborate mechanical design configurations needed for its implementation.

A parallel direction to the various TMD approaches is the concept of introducing negative stiffness elements (or 'anti-springs') for vibration isolation. This concept has also a long history, being first introduced in the pioneering publication of [33], as well as in the milestone developments of [35]. The central concept of these approaches is to significantly reduce the stiffness of the isolator and consequently to reduce the natural frequency of the system even at almost zero levels [6], being thus called "Quazi Zero Stiffness" (QZS) oscillators. In this way, the transmissibility of the system for all operating frequencies above the natural frequency is reduced, resulting to enhanced vibration isolation. An initial comprehensive review of such designs can be found in [21].

The negative stiffness behavior is primarily achieved by special mechanical designs involving conventional positive stiffness pre-stressed elastic mechanical elements, such as post-buckled beams, plates, shells and pre-compressed springs, arranged in appropriate geometrical configurations. Some interesting designs are described in [48], [44]. However, alternatively to elastic forces, other forms of physical forces can be used to produce an equivalent negative stiffness effect, such as gravitational [19], magnetic [36] or electromagnetic [52]. Quazi Zero Stiffness (QZS) oscillators are finding numerous applications in seismic isolation [17], [22], [37], [3] in all types of automotive suspensions [26], [27], [28] or in torsional vibrations [51]. Quite recently, periodic cellular structures with advanced dynamic behavior have been also proposed [45], [4], [32], [14], combining high positive and negative stiffness. Although the physical mechanisms that generate increased damping in cellular structures are not well understood, micro-buckling or slip-stick phenomena [25], [40], [11] could be among the possible explanations.

Parallel, quite interesting possibilities towards achieving significant damping have been demonstrated to exist also in materials comprising a negative stiffness phase [24], not only at a material level [41], but also at macroscopic devices [18]. Moreover, such a behaviour can be combined with high stiffness properties. A theoretical approach has been performed for the analysis of the static and dynamic stability of composite materials, incorporating negative stiffness elements [49].

However, Quazi Zero Stiffness (QZS) oscillators suffer from their fundamental requirement for a drastic reduction of the stiffness of the structure almost to negligible levels, which limits the static load capacity of such structures.

For the completeness of the review, it should be mentioned that a rich variety of nonlinear dynamic phenomena [43], [43], [27], [7], [23], [38], either inherently present, or designed to be present in all types of the above vibration absorbers, greatly contribute to the complexity of their dynamic behavior, as well as to the increase of their dynamic performance. However, their treatment is far beyond the scope of this paper. In any case, they can be used to act complimentary to the above vibration absorption concept.

Quite recently, a novel type of oscillator has been proposed [1], incorporating a negative stiffness element, which can exhibit extraordinary damping properties, without presenting the drawbacks of the traditional linear oscillator, or of the 'zero-stiffness' designs. This oscillator is designed to present the same overall (static) stiffness as a traditional reference original oscillator. However, it differs both from the original SDoF oscillator, as well as from the known negative stiffness oscillators, by appropriately redistributing the individual stiffness elements and by reallocating the damping. Although the proposed oscillator incorporates a negative stiffness element, it is designed to be both statically and dynamically stable. Once such a system is designed according to the approach proposed in [1], it is shown to exhibit an extraordinary damping behaviour. Moreover, a drastic increase of several orders of magnitude has been observed for the damping ratio of the flexural waves propagating within layered periodic structures incorporating such negative stiffness oscillators [12].

In this paper, the concept of [1] is treated in a systematic way, within the context of the design of a general class of tuned mass dampers. In this way, a direct comparison of the KDamper with the TMD is performed in section 2, which reveals its basic properties. The KDamper always indicates better isolation properties than a TMD damper with the same additional mass. Instead of increasing the additional mass, the vibration isolation capability of the KDamper can be increased by increasing the value of the negative stiffness element. Consequently, significant vibration isolation properties can be achieved, even for very low values (practically insignificant) of the additional mass. However, the increase of the negative stiffness element is upper bounded by the static stability limit of the structure.

Section 3 presents an application of the KDamper concept. The linear negative stiffness element is realized by a non-linear bistable element, which operates around an unstable equilibrium point. This bistable element takes the form of one Belleville spring, which transfer a negative force to a vertical spring through an appropriate mechanism. The system is designed so that it presents an adequate static load bearing capacity, while the transfer function of the system is below unity in the entire frequency range.

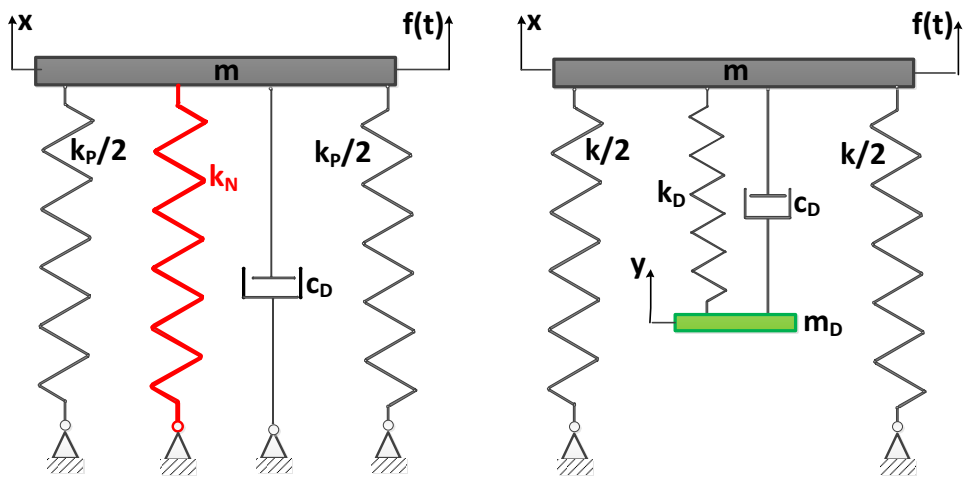

(a)            (b)

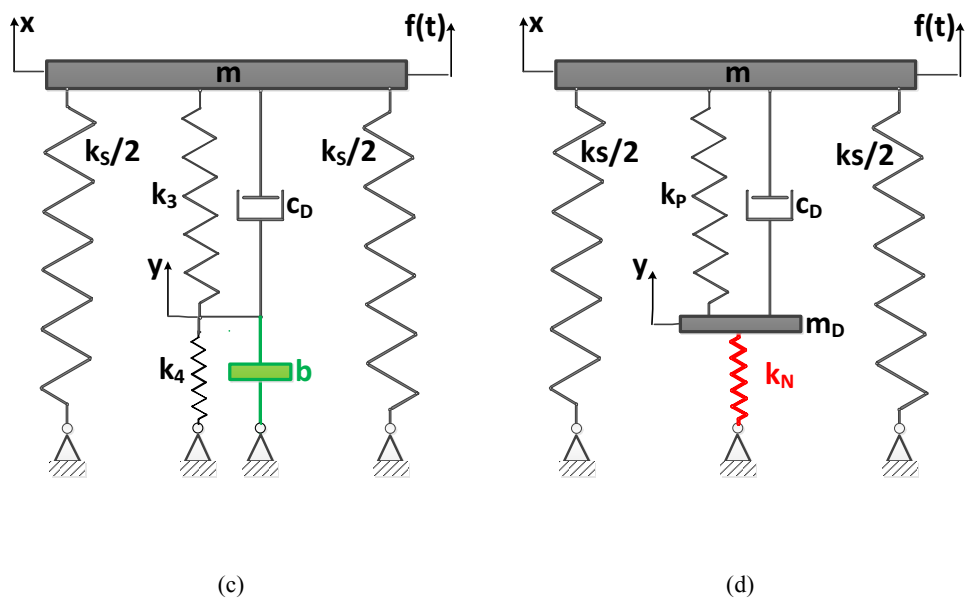

(c)                  (d)

**Figure 1:** *Schematic presentation of the considered vibration absorption concepts (a) Quasi-Zero Stiffness (QZS) oscillator, (b) Tunned Mass Damper(TMD), (c) Inerter (JDamper), (d) KDamper.*

Further insight to the physical behavior of the system is provided, by analyzing the balance of forces in the entire frequency range. It is observed that the magnitude of the damping force and of the inertia force of the additional mass is minimal throughout the entire frequency range, including the natural frequencies of the system. Thus, the external force is almost entirely balanced by the positive and negative stiffness forces, as well as by the inertia of the main mass. This fact ensures that an adequate level of elastic forces exists throughout the entire frequency range, able to counteract the inertial and the external excitation forces and justifies the concept of stiffness based vibration isolation.

As it has been mentioned the non-linear bistable element takes the form of one disc spring. Disc springs are conical ring washers whose shape changes under axial loads, based on the approximated rotation of the generally uniform rectangular cross-section of the disc around a circle of inversion. Compared to other types of springs, the disc spring can be categorized as having a "small spring deflection coupled with high spring force". However, this restriction is circumvented by the ability to form stacks of multiple disc springs. Arranging the discs in parallel or nested formation multiplies the spring force, alternating or series arrangement multiplies the spring deflection. Both these stacking methods can be used in combination.

One of the outstanding characteristics of the disc spring is doubtless its capacity for variation of the characteristic force-deflection curve over a wide range. Alongside practically linear characteristics, degressive force-deflection characteristics can also be implemented, even those in which spring force diminishes in certain ranges with increasing spring deflection. Many disc springs feature contact surfaces. These are predominantly large parts which in any case involve a high degree of production complexity. In this case, modified calculation methods are used. Contact surfaces improve the guidance properties of disc springs. In some applications, the guiding element of the disc spring stack can have a disturbing influence. Slotted disc springs assume a special role. The slotting process changes the force-deflection range of the individual disc springs, resulting in greater spring deflections coupled with lower spring force.

## 2. Basic properties of the KDamper

As it is stated in [2] KDamper uses the negative stiffness element and the basic concept as well as the steps of the optimal design are described.

Now the three basic properties of the KDamper are going to be stated. The first one is the following.

> **PROPERTY 1:** *The amplitude of the transfer function $T_{XKI}$ of the KDamper at the points $q_L$ and $q_R$ is less than the maximum amplitude of the transfer function $T_{XMmax}$ of a TMD with equal μ:*
>
> $$T_{XKI} = T_{XKL} = T_{XKR} \leq T_{XM\max} \tag{1}$$

The obvious consequence of PROPERTY 1 is that the addition of a negative stiffness spring reduces the magnitude of the transfer function of the TMD. Figure 4 presents the Transfer function of the KDamper for two values of κ.

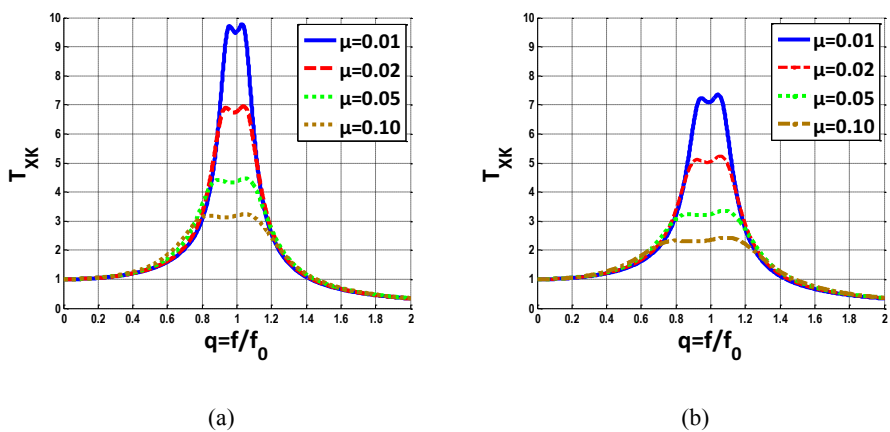

(a)          (b)

**Figure 4:** *Effect of the mass ratio μ on the Transfer function $T_{XK}$ of the KDamper for (a)κ=0.5, (b)κ=1.0.*

It should be noted that PROPERTY 1 does not hold for a spring $k_N$ with positive stiffness (i.e. with a negative value of κ).

Next, the property 2 is the following

> **PROPERTY 2:** *The amplitude of the transfer function $T_{XKI}$ of the KDamper at the points $q_L$ and $q_R$ tends to zero when $\kappa$ reaches the limit value of $\kappa_{MAX}$*
>
> $$T_{XKI} = T_{XKL} = T_{XKR} \to 0 \text{ for } \kappa \to \kappa_{MAX} \qquad (2)$$

The most important consequence of eq(2) is that quite small values of the transfer function $T_{XK}$ can be reached.

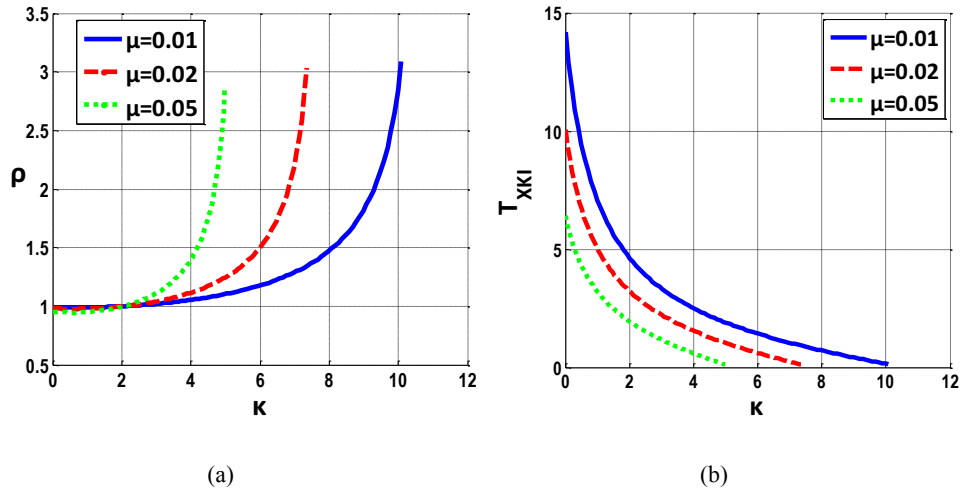

**Figure 5:** *Variation of the KDamper parameters a) Effect of $\kappa$ and $\mu$ the value of $\rho=\omega_D/\omega_0$. B) Effect of $\kappa$ and $\mu$ on the value $T_{XKI}$ of the transfer function $T_{XK}$ at the invariant points $q_L$, $q_R$.*

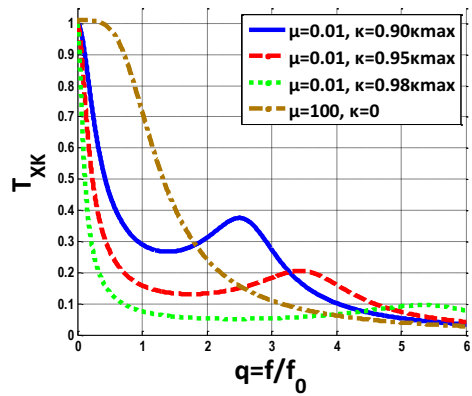

**Figure 6:** *Transfer functions $T_{XK}$ of the KDamper for values of $\kappa$, $\mu$ close to the limits.*

Figure 5 presents the variation of $\rho$ and $T_{XKI} = T_{XKL} = T_{XKR}$ with the increase of the parameter $\kappa$. Figure 6 presents Transfer functions for values of $\kappa$ or $\mu$ close to the limits. Values of low $\mu$/high $\kappa$ characterize the Kdamper, while values high $\mu$/low $\kappa$ can be indicative of an inerter. As it can be observed, very low values of the transfer function of the KDamper, can be reached, quite below unity. Moreover, these values can be achieved by an almost marginal value of $\mu=0.01$. This fact implies that the KDamper essentially does not require an additional mass $m_D$. Moreover, the Kdamper indicates a superior behavior at the very low frequency range.

A first implication of eq (42) is that there exists a range of values of $\kappa$ for which:

$$T_{XKL}(\kappa) = T_{XKL}(\kappa) < 1 \quad \text{for} \quad \kappa_{MAX} \geq \kappa \geq \kappa_{UN} \tag{3}$$

Eq. (43) implies that in this case, the values $T_{XKI}=T_{XKL}=T_{XKR}$ do no longer present the maximum values of $T_{XK}$ since $T_{XK}(q=0)=1$. Among others, this fact complicates the procedures for the selection of $\zeta_D$ based on averaging the slopes of $T_{XK}$ at the frequencies $q_L$, $q_R$, as it is the classical approach, proposed in [16].

Furthermore, although the fact that by increasing $\kappa$ the transfer function $T_{XKI}=T_{XKL}=T_{XKR}$ can be reduced almost to zero, increasing $\kappa$ has a number of implications in the design of the KDamper. First, high stiffness values result. Figure 7 reflects this fact to the increase of the stiffness values $k_N$, $k_P$ and especially $k_S$.

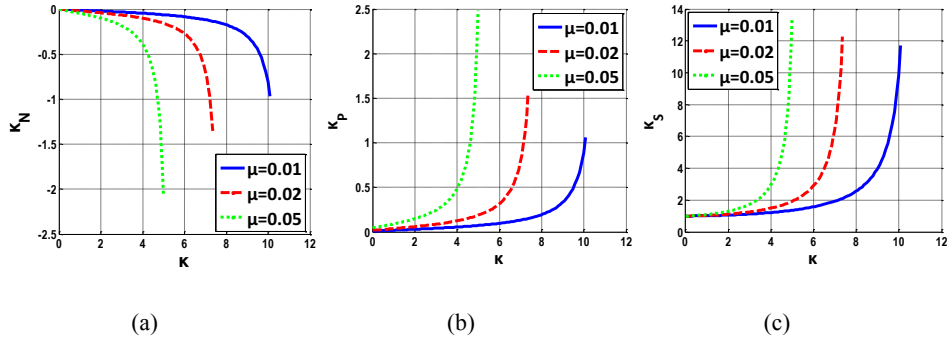

(a)     (b)     (c)

**Figure 7:** *Increase of the values of the stiffness elements of the KDamper by the increase of κ. (a) $\kappa_N$, (b) $\kappa_P$, (c) $\kappa_S$.*

Moreover, increasing the stiffness and especially $k_N$ may endanger the static stability of the structure. Although theoretically the value of $k_N$ is selected to ensure the static stability, variations of $k_N$ result in practice due to various reasons, such as temperature variations, manufacturing tolerances, or non-linear behavior, since almost all negative stiffness designs result from unstable non-linear systems. Consequently, an increase of the absolute value of $k_N$ by a factor ε may lead to a new value of $k_{NL}$ where the structure becomes unstable:

$$k_S + \frac{k_P k_{NL}}{k_P + k_{NL}} = 0 \Leftrightarrow k_{NL} = -\frac{k_S k_P}{k_S + k_P} = (1+\varepsilon)k_N \tag{4}$$

And so the static stability margin ε can be exressed as

$$\varepsilon = \frac{1}{\kappa[(1+(1+\kappa)^2 \mu \rho^2]} \tag{5}$$

As it can be shown, the following PROPERTY 3 holds for $\varepsilon$.

**PROPERTY 3:** The increase of the negative stiffness of the system is upper bounded by the static stability limit of the structure:

$$\varepsilon \to 0 \quad \text{for} \quad \kappa \to \kappa_{MAX} \tag{6}$$

Figure 8 presents the variation of $\kappa_{MAX}$ over $\mu$ and of $\varepsilon$ over $\kappa$ and $\mu$.

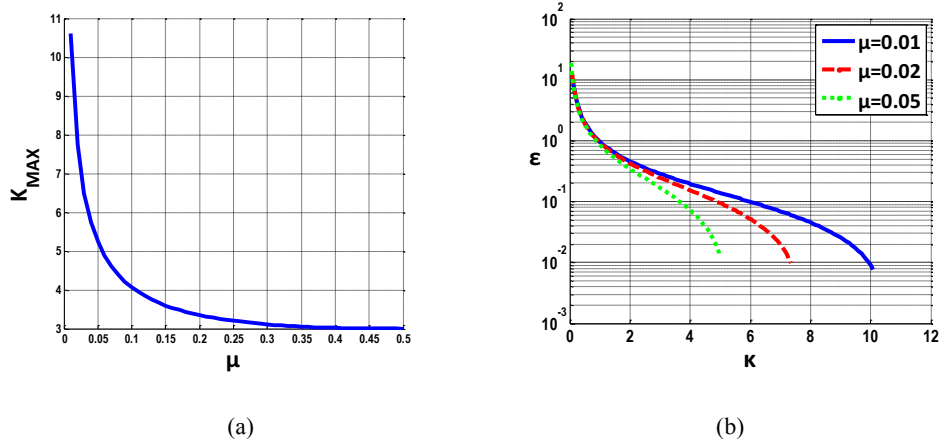

(a)  (b)

**Figure 8:** *Variation of KDamper parameters a) Effect of the mass ratio $\mu$ on the maximum value $\kappa_{MAX}$ of $\kappa$. b) Effect of $\kappa$ and $\mu$ over the static stability margin $\varepsilon$.*

Finally, high values of $\kappa$ result to increased amplitudes of the response $y$, which may encounter further technological constraints. For example, for $q=0$

$$T_{YXK}(q=0) = \frac{Y(q=0)}{X(q=0)} = \frac{Y(q=0)}{X_{ST}} = 1 + \kappa \tag{7}$$

## 3. An implementation example

### 3.1 Indicative realization of a negative stiffness spring

An example for an implementation of the KDamper is now considered. It consists from a mass $m$ which is supported by two parallel linear springs with stiffness $k_S$ and $k_P$ respectively and by a damper with constant $c_D$. The damper $c_D$ and the spring $k_P$ are also connected to a mass $m_D$. The negative stiffness spring $k_N$ is realized by one Belleville spring the properties of which will be found.

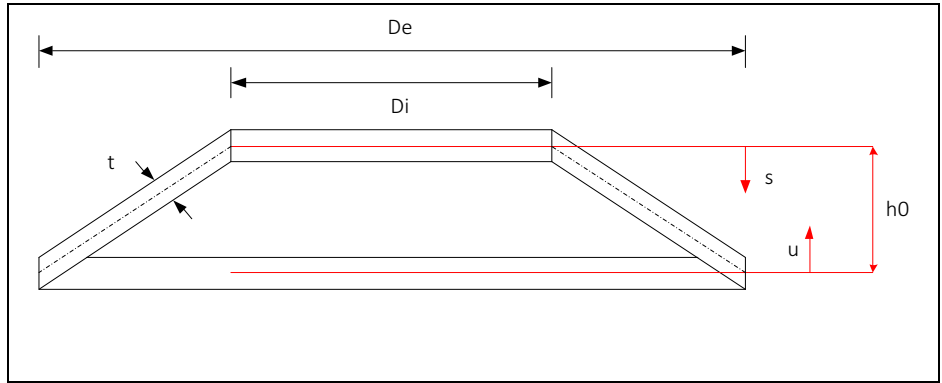

**Figure 9:** *Belleville spring*

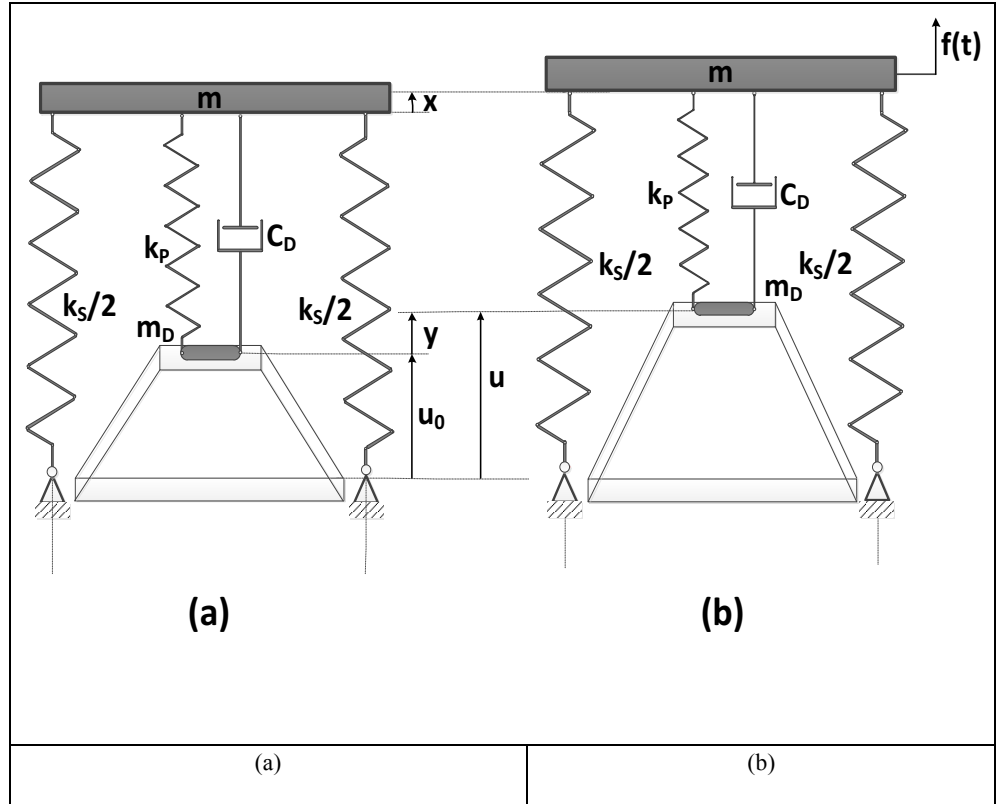

**Figure 10:** *Schematic presentation of the realization of the KDdamping concept using a Belleville spring (a) Configuration at the static equilibrium point, (b) Notation concerning the perturbed configuration.*

The static equilibrium position of the system is depicted in Fig 10(a), under the action of the gravity force. The perturbed position after an external dynamic excitation *f(t)* is depicted in Fig 10(b), along with the necessary notation concerning the various displacements of the system.

The equations of motion of the proposed oscillator are:

$$m\ddot{x} + c_D(\dot{x} - \dot{y}) + k_S(l_S - l_{SI}) + k_P(l_P - l_{PI}) = f + mg \quad (8.a)$$

$$m_D\ddot{y} - c_D(\dot{x} - \dot{y}) - k_P(l_P - l_{PI}) + f_N(u) = m_D g \quad (8.b)$$

4where $l_S(t)$ is the length of the spring $k_S$, $l_{SI}$ is the initial length of the un-deformed spring $k_S$, $l_P(t)$ is the length of the spring $k_P$, $l_{PI}$ is the initial length of the un-deformed spring $k_P$ and $f_N(u)$ is the non-linear force exerted by the Belleville spring.

The equations of the system at the static equilibrium point are derived by the equations (8):

$$k_S(l_{S0} - l_{SI}) + k_P(l_{P0} - l_{PI}) = mg \tag{9.a}$$

$$-k_P(l_{P0} - l_{PI}) + f_N(u_0) = m_D g \tag{9.b}$$

where the index $(\ )_0$ is used to denote the static equilibrium point, $l_{S0}$ is the length of the (normally deformed) spring $k_S$ at the static equilibrium point, $l_{P0}$ is the length of the (normally deformed) spring $k_P$ at the static equilibrium point and $f_N(u_0)$ is the non-linear force exerted by the Belleville spring at the static equilibrium point.

Denoting by:

$$v_S = l_{S0} - l_{SI} \tag{10.a}$$

$$v_P = l_{P0} - l_{PI} \tag{10.b}$$

$$f_{N0} = f_N(u_0) \tag{10.c}$$

the following equations result:

$$v_S = [(m + m_D)g - f_{N0}]/k_S \tag{11.a}$$

$$v_E = (-m_D g + f_{N0})/k_P \tag{11.b}$$

Further elaboration of the sets of Eqs (49),(50),(51) and substitution in the set of eqs (48) leads to the final set of equations of motion:

$$m\ddot{x} + c_D(\dot{x} - \dot{y}) + k_S x + k_P(x - y) = f \tag{12.a}$$

$$m_D \ddot{y} - c_D(\dot{x} - \dot{y}) - k_P(x - y) + f_{ND}(u) = 0 \tag{12.b}$$

where:

$$f_{ND}(u) = f_N(u) - f_{N0} \tag{13.a}$$

$$u = u_0 + y \tag{13.b}$$

$$l_S = l_{S0} + x \tag{13.c}$$

$$l_P = l_{P0} + x - y \tag{13.d}$$

The following expressions can be derived for the non-linear force $f_N$ and the equivalent non-liner stiffness $k_N$ of the Belleville spring (Lingshuai et al., 2014)

$$F = \frac{4E}{1-v^2} \frac{t^4}{K_1 D_e^2} K_4^2 \frac{s}{t} \left( K_4^2 \left(\frac{h_o}{t} - \frac{s}{t}\right)\left(\frac{h_o}{t} - \frac{s}{2t}\right) + 1 \right) \tag{14}$$

$$k_N = \frac{dF}{ds} = \frac{4E}{1-v^2} \frac{t^3}{K_1 D_e^2} K_4^2 \left(K_4^2 \left(\left(\frac{h_o}{t}\right)^2 - 3 \frac{h_o}{t} \frac{s}{t} + \frac{3}{2}\left(\frac{s}{t}\right)^2\right)\right) \quad (15)$$

**3.2 Selection of the system parameters**

The vibrations of a mass of m=100Kg are to be isolated, which is supported in the vertical direction by a system of springs of with a total static stiffness of $k$=49050N/m. The corresponding vertical static deflection of the system under its own weight is $X_{VSD}$=2cm and the natural frequency of the system is $f_0$=3.5Hz.

The main parameters of a KDamper system are selected as $\mu$=0.01 and $\kappa$=7.8. The resulting static stability margin is $\varepsilon$=0.05. The full set of the KDamper parameters is presented in Tables 1 and 2.

**Table 1.** Non-dimensional KDamper parameters

| $\mu$ | $\kappa$ | E | $\rho$ | $\zeta_D$ | $q_L$ | $q_R$ | $\kappa_N$ | $\kappa_P$ | $\kappa_S$ |
|---|---|---|---|---|---|---|---|---|---|
| 0.01 | 7.8 | 0.05 | 1.4279 | 0.69 | 0.97 | 1.87 | -0.159 | 0.1784 | 2.400 |

**Table 2.** Dimensional KDamper parameters

| $k_S$ | $k_P$ | $k_N$ | $m_D$ | $c_D$ |
|---|---|---|---|---|
| 1.17x10$^5$ N/m | 8.8004x10$^4$ N/m | -7.8003x10$^3$ N/m | 1.0 Kg | 43.64 Nsec/m |

It should be noted that if the damping element $C_D$ was used to connect directly the mass m to the support (ground), the equivalent damping effect on the system would have been essentially negligible:

$$\zeta_0 = c_D / 2\sqrt{km} = \zeta_D \rho \mu = 0.01 \quad (16)$$

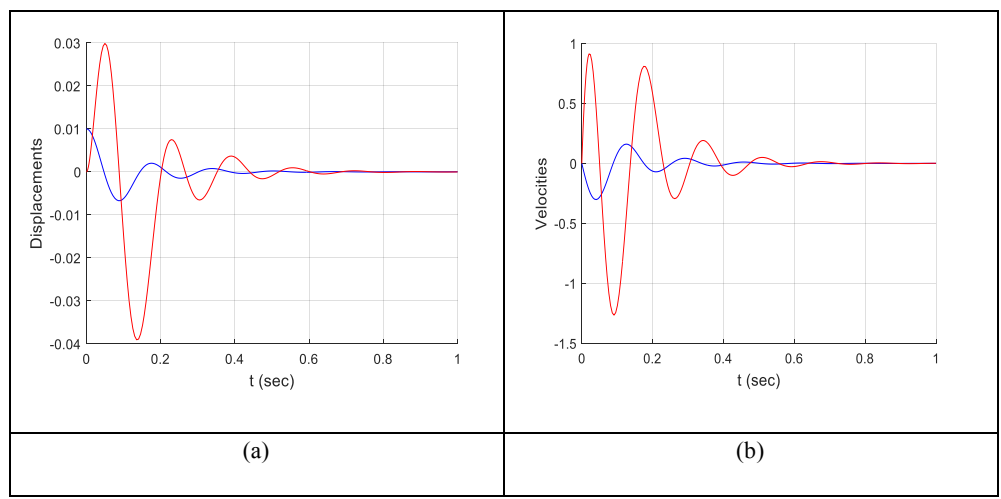

(a)      (b)

**Figure 11:** *Response Functions of the linear system to an initial displacement of 0.01 m a) Displacements, b) Velocities.*

Figure 11 presents the waveforms of the displacements and the velocities of the linear system to an initial displacement of 0.01m. As it can be observed at the figure above the maximum displacement of the mass $m_D$ is equal to $y_{max}$=0.04m.

The next step is the design of the Belleville springs to be used for the non-linear problem. First, the equation above for the nonlinear stiffness K is considered.

By setting the derivative of this equation equal to zero it can be found that when $k_N$ is maximum ($k_{max}$) s=$h_0$.

The following equations must be true

$$k_{max} = 1.02 k_N(\text{linear}) = -7946.7 \text{ N/m} \tag{17}$$

$$k(y_{max}) = 0.9 k_{max} = -7107.3 \text{ N/m} \tag{18}$$

The two equations must be true in order to have an adequate level of damping and also to keep the values of $k_N$ in the accepted limits. Therefore, from eq. (17),(18) the full set of the Belleville springs parameters can be defined. The full set of the Belleville springs parameters can be found in Table 3.

**Table 3.** Belleville springs parameters

| No | E | $v$ | $D_e$ | $D_i$ | th | $h_0$/th | d |
|---|---|---|---|---|---|---|---|
| 56 | 206 | 0.3 | 360 | 288 | 3 | 1.9 | 1.22 |

### 3.3 Response analysis for initial displacement

Figure 12 presents the waveforms of the displacements and velocities of the response of the non-linear oscillator to an initial displacement of 0.01m.

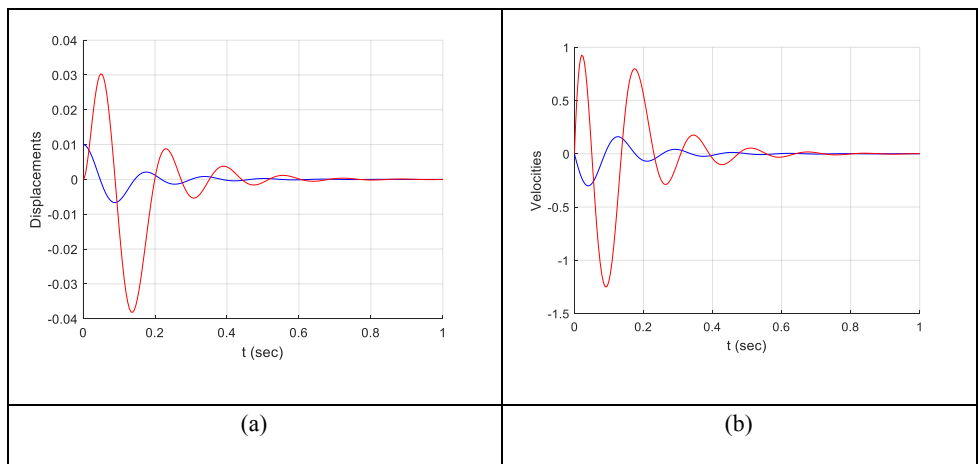

**Figure 12:** *Response of the KDamper configuration to an initial displacement of 0.01m (a) Displacements, (b) Velocities.*

Figure 13 presents the variation of the force $F_N$ as a function of the displacement $u$ of the Belleville springs. The red color represents the theoretical curve and the blue the real one.

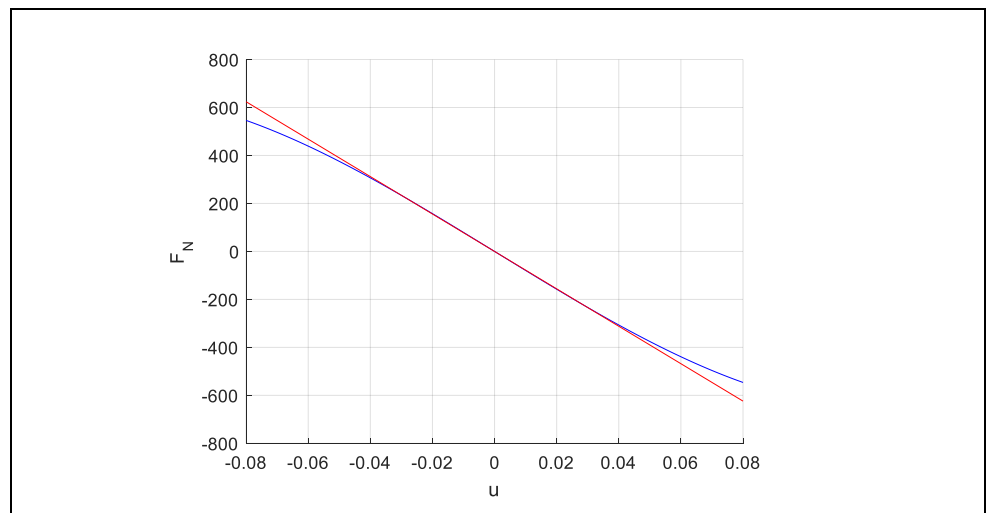

**Figure 13:** *Variation of the force $F_N$ as a function of the displacement u*

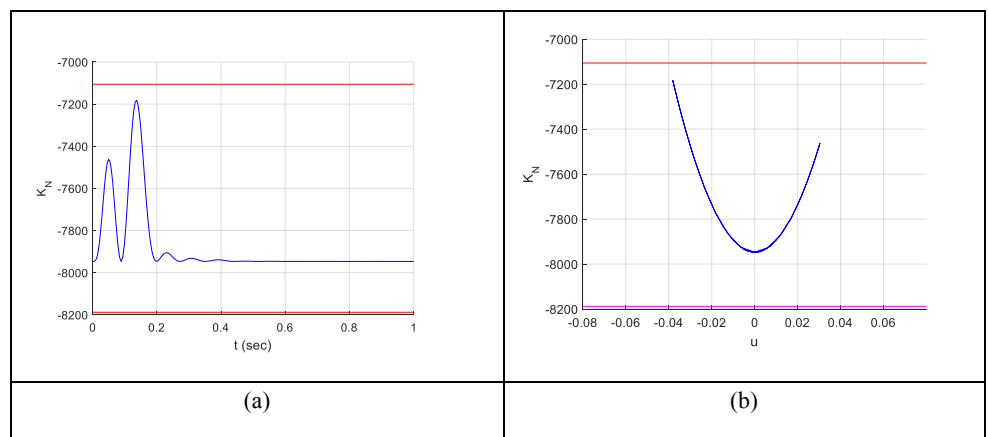

**Figure 14:** *Variation of the non-linear stiffness $k_N$ to an initial displacement 0.01m. a) Variation in the time domain, b) Range of oscillation.*

Figure 14a presents the variation of $k_N$ over time and Fig 14.b as a function of the displacement $u$ of the Belleville springs. A significant variation of $k_N$ is observed, verifying the strong no-linear nature of the response. However, the negative stiffness $k_N$ remains within the specified acceptable limits, which guarantee both static stability and damping behaviour.

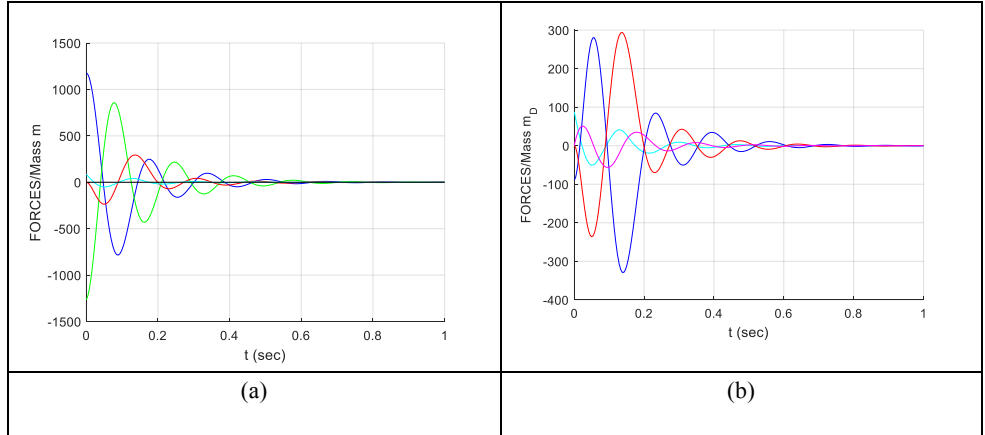

**Figure 15:** *Waveforms of the various force types of the KDamper to an initial displacement 0.01m (a) Forces at the mass m (b) Forces at the mass $m_D$*

Figure 15 presents the waveforms of the forces. This fact ensures that an adequate level of elastic forces exists throughout the entire frequency range, able to counteract the inertial and the external excitation forces and provides further insight on the background of stiffness based vibration isolation. The next step is to check the stresses for the proposed structure. The equations for the stresses in the Belleville springs are

$$\sigma_{OM} = -\frac{4E}{1-v^2} \frac{t^2}{K_1 D_e^2} K_4^2 \frac{s}{t} \frac{3}{\pi} \tag{19.a}$$

$$\sigma_I = -\frac{4E}{1-v^2} \frac{t^2}{K_1 D_e^2} K_4^2 \frac{s}{t} \left(K_4 K_2 \left(\frac{h_o}{t} - \frac{s}{2t}\right) + K_3\right) \tag{19.b}$$

$$\sigma_{II} = -\frac{4E}{1-v^2} \frac{t^2}{K_1 D_e^2} K_4^2 \frac{s}{t} \left(K_4 K_2 \left(\frac{h_o}{t} - \frac{s}{2t}\right) - K_3\right) \tag{19.c}$$

$$\sigma_{III} = -\frac{4E}{1-v^2} \frac{t^2}{K_1 D_e^2} K_4^2 \frac{s}{t} \frac{1}{\delta} (K_4(K_2 - 2K_3) \left(\frac{h_o}{t} - \frac{s}{2t}\right) - K_3) \tag{19.d}$$

$$\sigma_{IV} = -\frac{4E}{1-v^2} \frac{t^2}{K_1 D_e^2} K_4^2 \frac{s}{t} \frac{1}{\delta} (K_4(K_2 - 2K_3) \left(\frac{h_o}{t} - \frac{s}{2t}\right) + K_3) \tag{19.e}$$

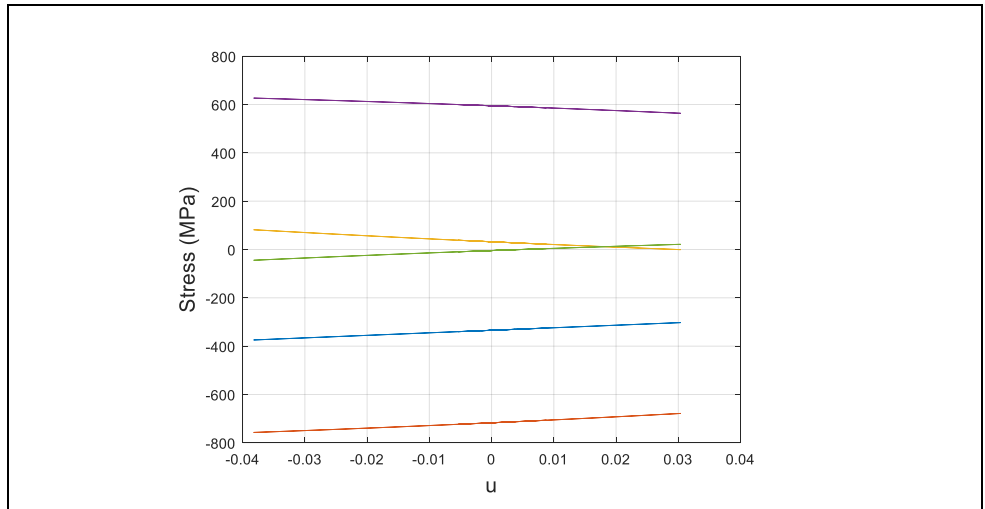

**Figure 16:** *Stresses in the Belleville spring for an initial displacement 0.01m*

As it can be seen in Fig.16 all stresses are in the allowed limits as the yield strength for the material proposed (steel) is 1100 MPa.

**3.4 Response analysis for initial velocity**

Figure 12 presents the waveforms of the displacements and velocities of the response of the non-linear oscillator to an initial velocity of 2.5m/s.

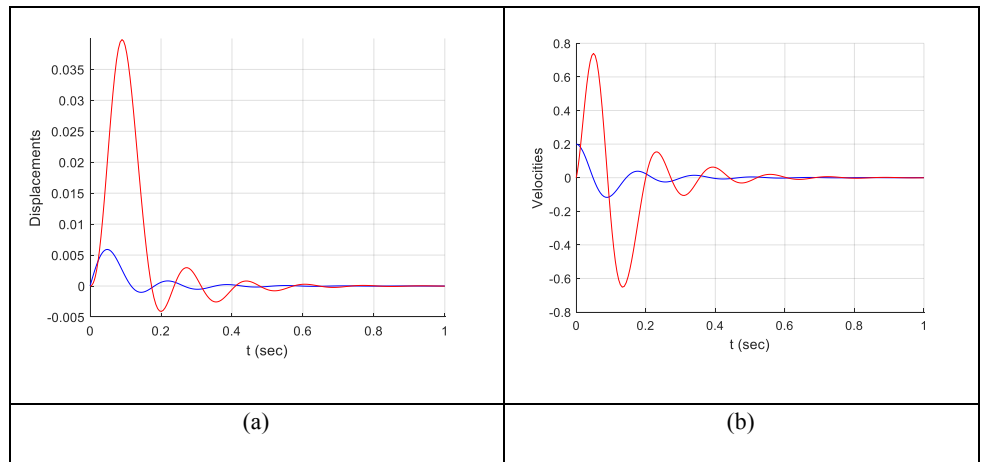

(a)            (b)

**Figure 12:** *Response of the KDamper configuration to an initial* velocity of 2.5m/s *(a) Displacements, (b) Velocities.*

Figure 13 presents the variation of the force $F_N$ as a function of the displacement $u$ of the Belleville springs. The red color represents the theoretical curve and the blue the real one.

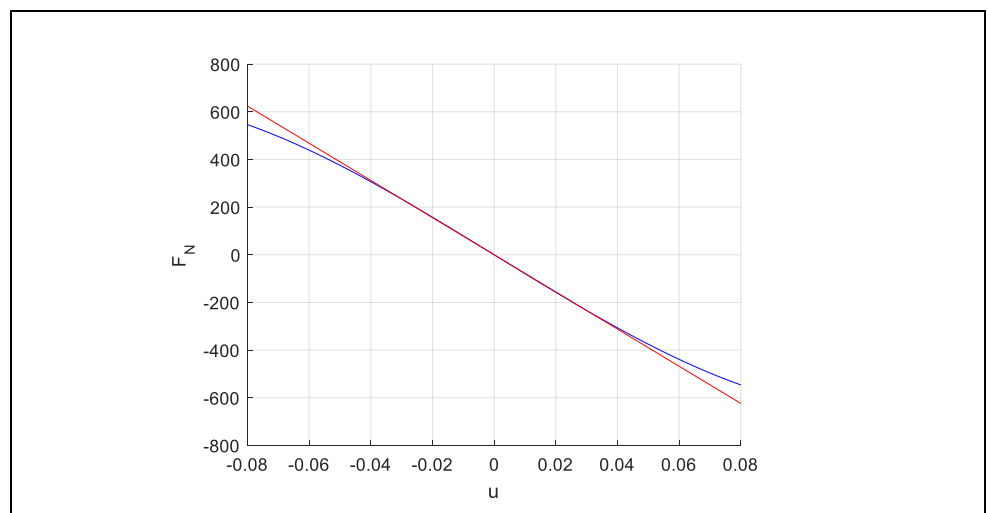

**Figure 13:** *Variation of the force $F_N$ as a function of the displacement u*

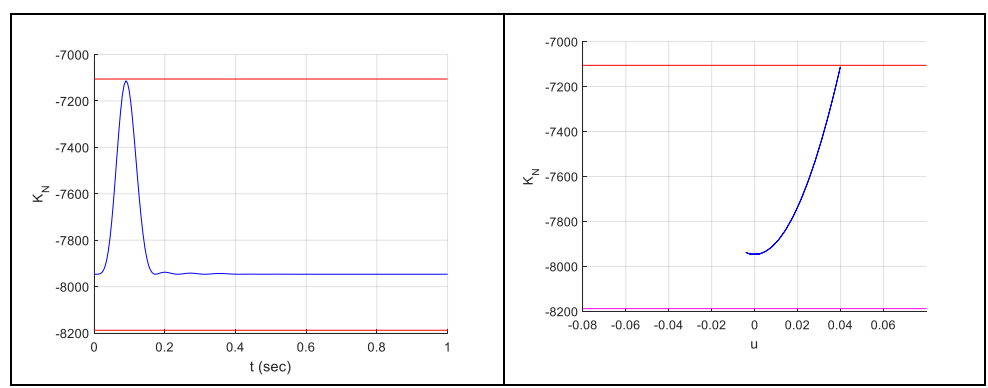

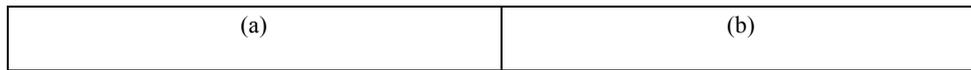

**Figure 14:** *Variation of the non-linear stiffness $k_N$ to an initial velocity of 2.5m/s. a) Variation in the time domain, b) Range of oscillation.*

Figure 14a presents the variation of $k_N$ over time and Fig 14.b as a function of the displacement $u$ of the Belleville springs. A significant variation of $k_N$ is observed, verifying the strong no-linear nature of the response. However, the negative stiffness $k_N$ remains within the specified acceptable limits, which guarantee both static stability and damping behaviour.

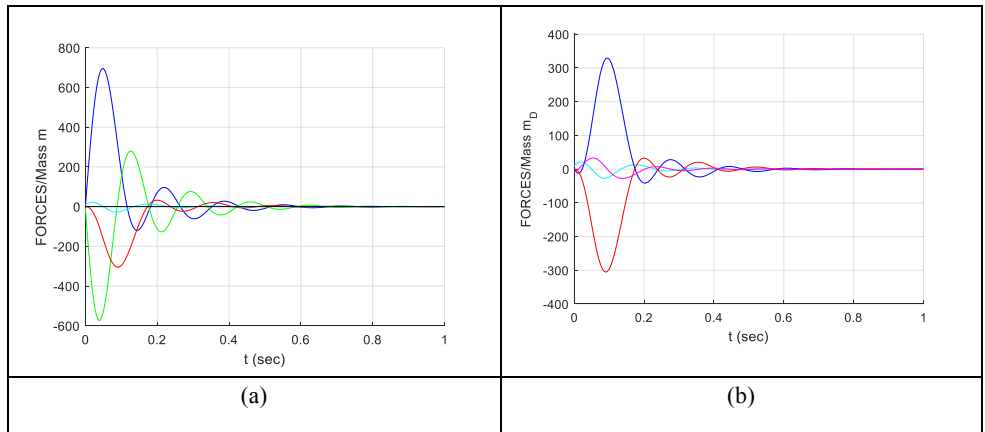

**Figure 15:** *Waveforms of the various force types of the KDamper to an initial velocity of 2.5m/s (a) Forces at the mass m (b) Forces at the mass $m_D$*

Figure 15 presents the waveforms of the forces. This fact ensures that an adequate level of elastic forces exists throughout the entire frequency range, able to counteract the inertial and the external excitation forces and provides further insight on the background of stiffness based vibration isolation.

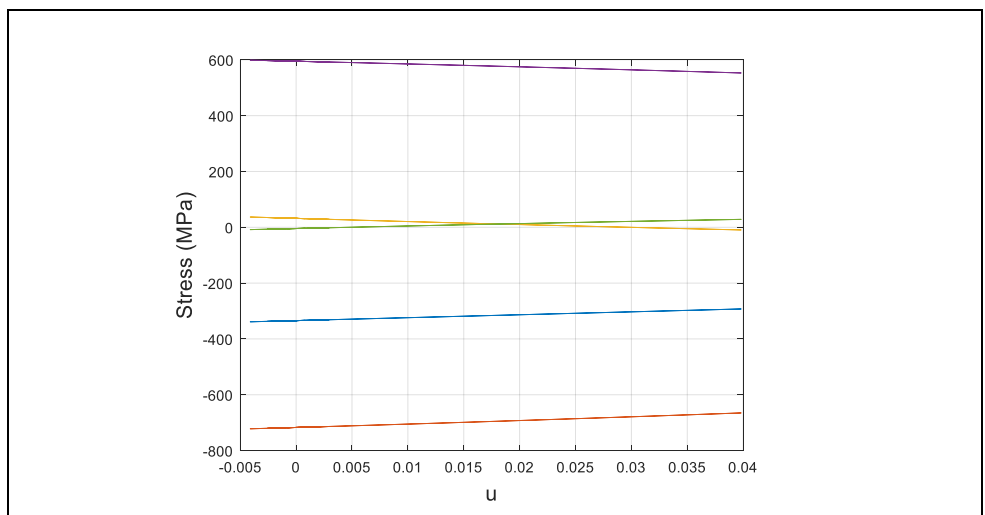

**Figure 16:** *Stresses in the Belleville spring for an initial displacement 0.01m*

As it can be seen in Fig.16 all stresses are in the allowed limits as the yield strength for the material proposed (steel) is 1100 MPa.

4.  **Conclusions**

The systematic design procedure of section 3 leads to a KDamper performance that can inherently offer far better isolation and damping properties than the TMD. Moreover, since the isolation and damping properties of the KD essentially result from the stiffness elements of the system, further technological advantages can emerge over other variants of TMD like the inerter, in terms of weight, complexity and reliability, without any need for compromises in the overall stiffness of the structure as it is the case of QZS.

The background of the performance of this stiffness based vibration absorption concept is the fact that an adequate level of elastic forces exists throughout the entire frequency range, able to counteract the inertial and the external excitation forces, while the damping forces and the inertia forces of the additional mass remain minimal in the entire frequency range, including the natural frequencies.

Moreover, the inherent non-linear nature of the negative stiffness force can be exploited to offer further potential advantages of the KDamper concept, such as robustness, broadband response and energy sinks.

It should be mentioned that the approach presented does not simply refer to discrete vibration absorption device, but it consists a general vibration absorption concept, applicable also for the design of advanced materials or complex structures. Such a concept thus presents the potential for numerous implementations in a large variety of technological applications, while further potential may emerge in a multi- physics environment.

**Aknowledgements**

The first author would like to express his sincere thanks to the "The Foundation for Education and European Culture" for the financial support he was granted for his PhD studies during of which this work was implemented